\documentclass[aps, prl, amsmath, amssymb, twocolumn, showpacs, superscriptaddress, reprint]{revtex4}

\usepackage{graphicx}
\usepackage{epstopdf}
\usepackage{bm}
\usepackage{longtable}
\usepackage{float}
\usepackage{dcolumn}
\usepackage{bm}

\begin{document}


\title{Ramsey Spectroscopy with Displaced Frequency Jumps}

\author{M. Shuker}
\email{moshe.shuker@nist.gov}
\affiliation{National Institute of Standards and Technology, Boulder, Colorado 80305, USA}
\affiliation{University of Colorado, Boulder, Colorado 80309-0440, USA}

\author{J. W. Pollock}
\affiliation{National Institute of Standards and Technology, Boulder, Colorado 80305, USA}
\affiliation{University of Colorado, Boulder, Colorado 80309-0440, USA}

\author{R. Boudot}
\affiliation{FEMTO-ST, CNRS, 26 rue de l'\'epitaphe 25030 Besancon, France}
\affiliation{National Institute of Standards and Technology, Boulder, Colorado 80305, USA}

\author{V. I. Yudin}
\affiliation{Novosibirsk State University, ul. Pirogova 2, Novosibirsk, 630090, Russia }
\affiliation{Institute of Laser Physics SB RAS, pr. Akademika Lavrent'eva 13/3, Novosibirsk, 630090, Russia }
\affiliation{Novosibirsk State Technical University, pr. Karla Marksa 20, Novosibirsk, 630073, Russia }

\author{A. V. Taichenachev}
\affiliation{Novosibirsk State University, ul. Pirogova 2, Novosibirsk, 630090, Russia }
\affiliation{Institute of Laser Physics SB RAS, pr. Akademika Lavrent'eva 13/3, Novosibirsk, 630090, Russia }

\author{J. Kitching}
\affiliation{National Institute of Standards and Technology, Boulder, Colorado 80305, USA}
\affiliation{University of Colorado, Boulder, Colorado 80309-0440, USA}

\author{E. A. Donley}
\affiliation{National Institute of Standards and Technology, Boulder, Colorado 80305, USA}
\affiliation{University of Colorado, Boulder, Colorado 80309-0440, USA}

\date{\today}


\begin{abstract}
Sophisticated Ramsey-based interrogation protocols using composite laser pulse sequences have been recently proposed to provide next-generation high-precision atomic clocks with a near perfect elimination of frequency shifts induced during the atom-probing field interaction. We propose here a simple alternative approach to the auto-balanced Ramsey (ABR) interrogation protocol and demonstrate its application to a cold-atom microwave clock based on coherent population trapping. The main originality of the method, based on two consecutive Ramsey sequences with different dark periods, is to sample the central Ramsey fringes with frequency-jumps finely-adjusted by an additional frequency-displacement concomitant parameter, scaling as the inverse of the dark-period. The advantage of this displaced frequency-jumps Ramsey (DFJR) method is that the local oscillator frequency is used as a single physical variable to control both servo loops of the sequence, simplifying its implementation and avoiding noise associated with controlling the LO phase. Compared to the usual Ramsey-CPT technique, the DFJR scheme reduces the sensitivity of the clock frequency to variations of the CPT sideband ratio and to the one-photon laser detuning by more than an order of magnitude. This simple method could be applied in a wide variety of Ramsey-spectroscopy based applications including frequency metrology with CPT-based and optical atomic clocks, mass spectrometry, and precision spectroscopy.
\end{abstract}

\maketitle

Ramsey's method of separated oscillating fields \cite{RamseyRhysRev1950} enables the measurement of atomic spectral lines with unrivaled precision and accuracy. This widely-used spectroscopy and interrogation technique has been successfully applied in a variety of quantum devices and instruments including atomic frequency standards \cite{ThomasPRL1982, SantarelliPRL1999, ZanonPRL2005, NicholsonNatComm2015, NemitzNatPhoton2016, SchioppoNatPhoton2017}, matter-wave interferometry and atomic inertial sensors \cite{RiehlePRL1991, MenoretSR2018}, Bose-Einstein condensates \cite{DonleyNature2002}, quantum computing \cite{LimSR2014}, memories \cite{MaitrePRL1997} and information processing \cite{SchindlerNJP2013}, cavity quantum electrodynamics \cite{DelegliseNature2008}, optomechanics \cite{QuPRA2014}, extreme-ultraviolet spectroscopy \cite{PirriPRA2008}, mass spectrometry \cite{BollenNIMB1992} as well as the experimental confirmation of fundamental quantum mechanics concepts \cite{GleyzesNature2007}.

In a typical Ramsey sequence, atoms are probed with two successive pulses separated by a free-evolution period (or dark period) $T$. The first pulse creates a coherence between targeted quantum states whose initial phase depends on the frequency and amplitude of the external field as well as on the pulse duration. During the free-evolution period, a phase shift develops between the atomic coherence and the excitation field determined by their frequency difference and $T$. Through an interference process, the second pulse leads to the modulation of the field absorption (or transmission) spectrum allowing for an effective measurement of the phase difference between the atomic precession and the local oscillator (LO) at the time of the second pulse. Scanning the frequency of the excitation field around the exact resonance and measuring the absorption (or transmission) of the second pulse leads to the detection of a pattern of Ramsey fringes, whose line-width scales as $1/(2T)$.

For clock applications, Ramsey spectroscopy reduces the sensitivity of the clock frequency to variations of the interrogating field \cite{RamseyRhysRev1950,HafizJAP2017}. However, the Ramsey interrogation technique exhibits a non-negligible residual sensitivity to frequency-shifts induced by the probe field during the preparation pulse. These light-shifts can be a key limiting factor to the accuracy and long-term fractional frequency stability of atomic clocks based on coherent population trapping (CPT) \cite{HafizJAP2017, YunPRappl2017, LiuPRAppl2017, LiuAPL2017} and optical clocks that probe ultra-narrow octupole \cite{HosakaPRA2009} or two-photon optical transitions \cite{FischerPRL2009, BadrPRA2006}.

In Ramsey-CPT clocks, light-shifts depend on the one-photon detuning of the laser, the total laser intensity, and the CPT optical sideband intensity ratio. The light-shifts originate from resonant interaction of the dual-frequency optical field with the 3-level atomic $\Lambda$-scheme \cite{HemmerJOSAB1989,BlanshanPRA2015} and from off-resonant coupling of the light field components with neighboring detuned atomic energy levels \cite{YanoPRA2014,PatiJOSAB2015}. The resonant shift, that scales inversely with the Ramsey period, results from incomplete dark-state formation during the first Ramsey-CPT pulse, and can be canceled by increasing the duration or intensity of the first Ramsey pulse \cite{HemmerJOSAB1989, ShahriarPRA1997, PatiJOSAB2015, LiuPRAppl2017}. The off-resonant shift, depending strongly on the CPT interrogation polarization scheme and the relative intensity of both CPT optical lines but weakly on the total intensity \cite{PollockPRA2018}, remains the dominant effect limiting the clock performance.

Over the last decade, sophisticated interrogation protocols based on Ramsey's method and composite laser pulses have been proposed in order to eliminate probe-induced frequency shifts \cite{ZanonRPP2018}. Among them, the auto-balanced Ramsey (ABR) scheme \cite{SannerPRL2018} is based on the extraction of two error signals derived from two successive Ramsey sequences with different dark periods. The first feedback loop uses the error signal generated by the short Ramsey sequence to apply a phase-step correction to the local oscillator during the dark period that nulls the probe-field induced frequency shift. The second loop stabilizes the LO frequency using the error signal derived from the long Ramsey sequence. The ABR technique has been applied to different kinds of atomic clocks, demonstrating a significant reduction of the clock frequency sensitivity to light-shifts and improving the clock mid and long-term frequency stability performance \cite{SannerPRL2018, HafizPRAppl2018, HafizAPL2018}. In \cite{YudinPRAppl2018} a theoretical generalization of the ABR method was presented, demonstrating the possibility to use various concomitant parameters including frequency steps during the Ramsey pulses or variations of the Ramsey pulses duration. 

In this letter, we propose and demonstrate a simple method to reduce interrogation-related frequency-shifts in Ramsey spectroscopy that we call Ramsey spectroscopy with displaced frequency jumps (DFJR). Similarly to the techniques presented previously \cite{SannerPRL2018,YudinPRAppl2018,HafizPRAppl2018}, two servo loops are used to control the clock frequency and a concomitant parameter that suppresses interrogation shifts. However, we propose a way to implement both control parameters with a single physical variable - the LO frequency. Aside from the requirement that the phase of the interrogation field is stable during the Ramsey cycle, no modulation or control of the LO phase is needed. This allows for simpler implementation and the elimination of the systematics associated with controlling and modulating the phase of the oscillator. We also note that the LO frequency remains constant throughout the Ramsey cycle (like in a standard Ramsey-CPT interrogation) and no rapid control of the LO frequency is required. This method can be applied to any Ramsey spectroscopy measurement, including for CPT atomic clocks and optical atomic clocks.

In a Ramsey clock, by sampling the half-height signal on both sides of the central Ramsey fringe with two consecutive Ramsey cycles, a zero-crossing error signal can be derived by subtraction of the two measurements and used to stabilize the frequency of the LO onto the fringe center. Two options are generally used for the generation of the error signal. One option, known as phase-jumps, is to  abruptly change the phase of the LO between the first and second Ramsey pulse by $\pi/2$ (-$\pi/2$) during the first (second) Ramsey cycle. Another option, known as frequency-jumps, is to jump the frequency of the LO to $1/(4T)$ ($-1/(4T)$) from the estimated clock frequency during the first (second) Ramsey cycle. 
\begin{figure}[t]
\centering
    \includegraphics[width=0.85\linewidth]{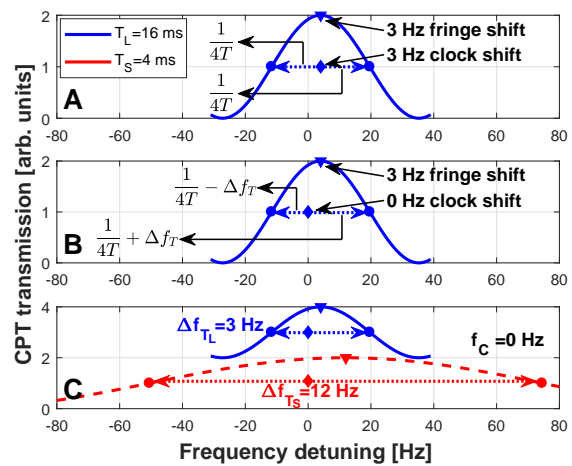}
    \caption{Illustration of light shifts and their mitigation using the DFJR method. The graphs show the central Ramsey-CPT fringes obtained for two different dark-periods ($T_L=$16 ms: blue line and $T_S=$4 ms: red dashed line). The center of each fringe is marked with a triangle. The steady-state value of the clock frequency is marked with a diamond symbol. Dotted lines show the frequency jumps applied to the sampling frequencies (marked with circles). A. Light shift with standard Ramsey-CPT interrogation ($T=16$ ms). The clock stabilizes to the shifted Ramsey fringe. B. By displacing the frequency jumps - jumping a different amount to higher and lower frequencies, the clock frequency shifts can be mitigated.  The challenge is then to find the suitable frequency displacement $\Delta f_T$. C. The DFJR method is used to find $\Delta f_T$ and stabilize the clock frequency on the non-shifted resonance frequency ($f_c=0$ Hz). Using two different dark-periods, $T_S=4$ ms (lower part of the graph) and $T_L=16$ ms (upper part of the graph) and adding the frequency displacement $\Delta f_T$ as a $T$-dependent control parameter, the clock frequency is free from light shifts.}
   \label{image:DFJ_Illustration}
\end{figure}

The basic concept of the DFJR method is illustrated in Fig. \ref{image:DFJ_Illustration}. Figure \ref{image:DFJ_Illustration}.A depicts a Ramsey-CPT fringe for $T=16$ ms, probed with standard $\pm 1/(4T)$ frequency-jump interrogation, and assumed to be light-shifted by $3$ Hz, which results in a similar shift of the stabilized clock frequency. This frequency shift can be eliminated by applying frequency-jumps of different magnitude to the right and left sides of the fringe such that the error signal steers the clock to the non-shifted frequency. This approach is demonstrated in Fig. \ref{image:DFJ_Illustration}.B where the interrogation-asymmetry is denoted by $\Delta f_T$ - the frequency displacement which depends on the dark period T: 
\begin{equation}\label{alpha_definition}
\Delta f_T=\frac{\alpha}{T}
\end{equation}
where $\alpha$ is the concomitant control parameter in the DFJR method. Without prior knowledge of the shifts in the system (which may also vary over time), the difficulty is then to find the correct $\Delta f_T$ value to apply. 

For this purpose, the DFJR method consists of a composite Ramsey sequence involving two different dark-periods, and use the measured error signals to find the suitable $\Delta f_T$ value and adapt it over time to compensate drifts in the light-shift. We take advantage of the fact that the interrogation-related frequency shifts depend inversely on the dark-period. By defining $\Delta f_T$ as a $T$-dependent parameter, the interrogation-related shifts can be distinguished and separated from the clock frequency $f_c$ (which is independent of T). Figure \ref{image:DFJ_Illustration}.C illustrates the principle of the DFJR method. Two Ramsey fringes obtained with a long dark period $T_L=16$ ms and a short dark period $T_S=4$ ms are assumed to be light-shifted by $3$ Hz and $12$ Hz respectively. The frequency jumps are displaced by $\Delta f_{T_L}=\frac{\alpha}{T_L}$ and $\Delta f_{T_S}=\frac{\alpha}{T_S}$ respectively. Two error signals are then generated (from the short and long cycles) and are used to stabilize the clock frequency $f_c$ and the concomitant control parameter $\alpha$. The only steady-state solution that nulls both error signals occurs when the clock frequency is on resonance, as depicted in Fig. \ref{image:DFJ_Illustration}.C. This results in a clock frequency free from the influence of interrogation related shifts.

\begin{figure}[t]
\centering
    \includegraphics[width=\linewidth]{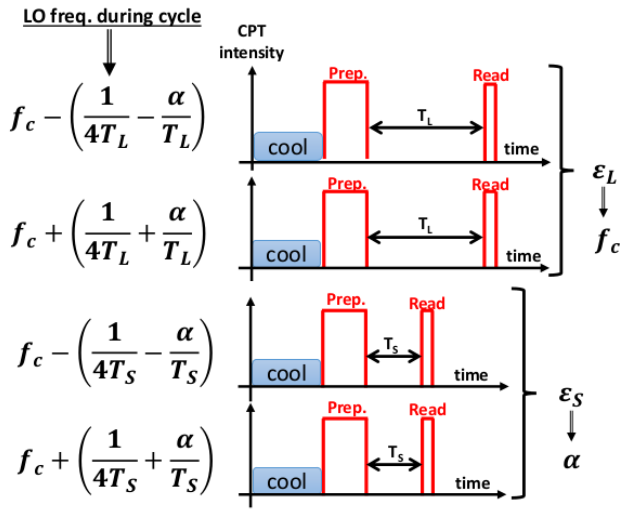}
    \caption{An illustration of the DFJR sequence, applied to a cold-atom CPT clock. The sequence is composed of four consecutive Ramsey cycles -- two with a long dark period, $T_L$ and two with a short dark period $T_S$. For each dark period, the two cycles include frequency jumps to the side of the central fringe. However, frequency jumps are corrected by a frequency displacement parameter $\Delta f_T$, which scales as the inverse of the dark period (see Eq. \ref{alpha_definition}). The frequency for each Ramsey cycle is noted on the left side of the figure. The error signal constructed from the long cycle-pair, $\varepsilon_L$, is used to control the clock frequency $f_c$. The error signal derived from the short cycles, $\varepsilon_S$, is used to control the concomitant control parameter $\alpha$.}
   \label{image:FJABRSequence}
\end{figure}

Figure \ref{image:FJABRSequence} shows the proposed DFJR sequence. It consists of two cycles with a long dark period $T_L$ and two cycles with a short dark period $T_S$. For each pair of cycles, the error signal is generated using frequency jumps from measurements of the half-height signal values on both sides of the central fringe (one value per cycle) and extracting their difference. In the long dark-period cycles, the sampling frequencies are $f_c-\left(\frac{1}{4T_L}-\frac{\alpha}{T_L}\right)$ and $f_c+\left(\frac{1}{4T_L}+\frac{\alpha}{T_L}\right)$, while for the short dark-period cycles the corresponding frequencies are $f_c-\left(\frac{1}{4T_S}-\frac{\alpha}{T_S}\right)$ and $f_c+\left(\frac{1}{4T_S}+\frac{\alpha}{T_S}\right)$. Thus in the DFJR method, one control parameter (the clock frequency, $f_c$) is the same for both dark-periods, while the other control parameter, $\alpha$, causes a frequency displacement that scales with the dark-period. This allows one to implement two control parameters combined in a single physical variable (the LO frequency). The error signal from the long dark-period cycle pair $\varepsilon_L$ is used to steer the clock frequency, while the error signal from the short dark-period cycle pair $\varepsilon_S$ is used to steer $\alpha$. In the presence of frequency shifts that depend inversely on the duration of the dark-period, the control system will stabilize such that the frequency displacement parameter adapts to compensate these shifts and the clock frequency is free from them. Note that the total DFJR sequence shown in Fig. \ref{image:FJABRSequence} is symmetrized as reported in \cite{HafizAPL2018} for optimal efficiency of the light-shift rejection.

We implemented the DFJR protocol utilizing a cold-atom atomic clock apparatus previously described in \cite{BlanshanPRA2015, LiuPRAppl2017}. A six-beam magneto-optical trap (MOT) is applied for $20$ ms followed by a $3$ ms molasses, traping and cooling $\sim 1\times 10^{6}$ $^{87}$Rb atoms at a temperature of about $\sim 10$ $\mu$K. At this stage, the atoms are  allowed to fall freely and are interrogated in a Ramsey-CPT cycle. A $3$ ms CPT preparation pulse is followed by a dark-period $T=4$-$16$ ms and a  $50$ $\mu$s reading CPT pulse. We use the lin $||$ lin CPT interrogation scheme \cite{TaichenachevJETP2005,ZibrovPRA2010} to enhance the CPT contrast and retro-reflect the CPT beam to minimize Doppler shifts \cite{EsnaultPRA2013}. Three sets of Helmholtz coils are used to set a quantization magnetic field of $B_z=4.4$ $\mu$T in the direction of the CPT beam propagation. The CPT beam is generated from a distributed Bragg reflector (DBR) laser modulated at $6.835$ GHz using an electro-optic modulator (EOM). The optical carrier and the $-1$-order sideband are used as the two CPT fields. The intensity ratio between both CPT fields is controlled using the EOM and measured by a Fabry-Perot interferometer. The intensity and frequency (one-photon detuning, OPD) of the CPT beam are controlled using an acousto-optic modulator (AOM). The Ramsey protocols are implemented by controlling the frequency synthesizer driving the EOM. The synthesizer is referenced to a hydrogen maser, and only the frequency of the synthesizer is controlled to implement the protocols. By stabilizing the synthesizer frequency according to the error signals, an atomic clock is obtained and the absolute frequency shifts are measured. 

\begin{figure}[t]
    \includegraphics[width=8.6cm]{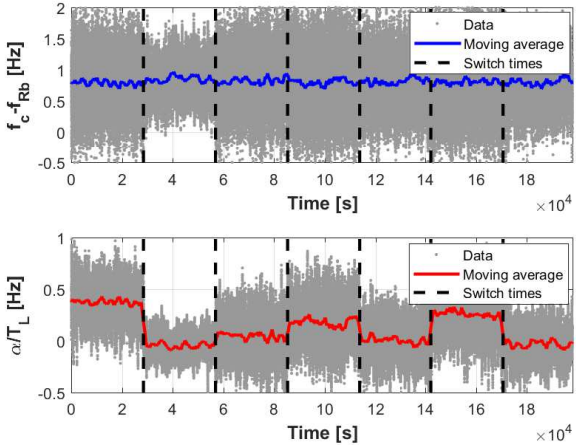}
    \caption{
    \label{image:ClockTraceSwitchRF}
    A time trace of the atomic clock operation based on DFJR. The upper pane shows the clock frequency shift $f_c-f_{Rb}$ and the lower pane shows the frequency displacement for the long cycle $\Delta f_{T_L}=\frac{\alpha}{T_L}$. Each pane shows the original data (dots) and a moving average (solid line). During the clock run the CPT intensity ratio is abruptly changed (switch times are shown in black dashed lines), changing the off-resonant light shift of the clock transition. It is evident that $\alpha$ adapts to accommodate the changing light shift leaving the clock frequency $f_c$ unaffected (at the value of the $2^{nd}$ order Zeeman shift). 
    }
\end{figure}

The DFJR method was compared to standard Ramsey-CPT spectroscopy by running the cold-atom clock under varying experimental parameters and interrogation shifts. Figure \ref{image:ClockTraceSwitchRF} depicts a time trace of DFJR clock operation. The upper pane shows the clock frequency shift, $f_c-f_{Rb}$, where $f_{Rb}$ is the generally accepted unperturbed $^{87}$Rb hyperfine frequency \cite{RiehleMetrologia2018}, while the lower pane shows the frequency displacement for the long cycle $\Delta f_{T_L}=\frac{\alpha}{T_L}$. During clock operation, the off-resonant light shift is altered by abruptly changing the CPT intensity ratio. Each time the CPT intensity ratio is changed, resulting in a different light shift, the frequency displacement control parameter $\alpha$ stabilizes to a new value, compensating for the light shift, and keeping the clock frequency $f_c$ unaffected. In essence, the two-loop control system keeps one of the control parameters (the clock frequency) protected from this particular kind of systematic. 

\begin{figure}[t]
    \includegraphics[width=8.6cm]{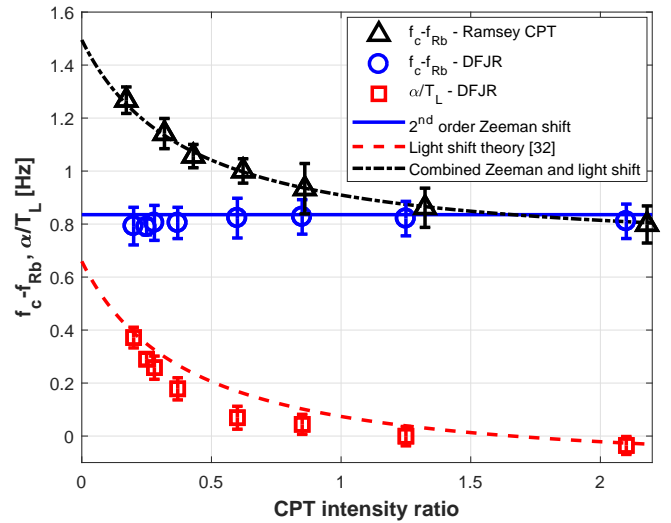}
    \caption{
    \label{image:ShiftVsRatio}
    Frequency shift $f_c-f_{Rb}$ of the clock, in Ramsey-CPT (triangles, with $T=16$ ms) and DFJR (circles, with $T_S=4$ ms and $T_L=16$ ms) methods, and frequency displacement for the long cycle $\alpha/{T_L}$ (squares) versus the CPT intensity ratio. In the Ramsey-CPT case, the clock frequency is shifted when the CPT intensity ratio is changed, due to off-resonant light-shift \cite{PollockPRA2018}. In the DFJR case, the clock frequency is nearly constant, and $\alpha$ changes to compensate for the light shift. The value of the single control parameter in the Ramsey-CPT scheme (the clock frequency) is effectively split into the two control parameters with the DFJR method. The frequency displacement $\Delta f_T$ is associated with the interrogation-related shifts. The clock frequency $f_c$ is free from interrogation related shifts, and is only shifted by the $2^{nd}$-order Zeeman shift.
    }
\end{figure}

Figure \ref{image:ShiftVsRatio} shows the clock frequency shift $f_c-f_{Rb}$ versus the CPT intensity ratio in the standard Ramsey-CPT regime ($T$ = 16 ms), compared to the DFJR scheme ($T_S=4$ ms and $T_L=16$ ms). In the standard Ramsey-CPT scheme (triangles) the clock frequency significantly changes with the CPT intensity ratio due to the off-resonant light shift \cite{PollockPRA2018}. With the DFJR scheme, the frequency displacement parameter $\alpha$ (squares) changes with the CPT intensity ratio in order to compensate for the light-shift, and the clock frequency (circles) variations are considerably reduced. The absolute clock frequency shift is then about 0.835 Hz, in good agreement with the expected second-order Zeeman shift in the lin $||$ lin configuration (which is T-independent and hence not canceled).

\begin{figure}[t]
    \includegraphics[width=8.6cm]{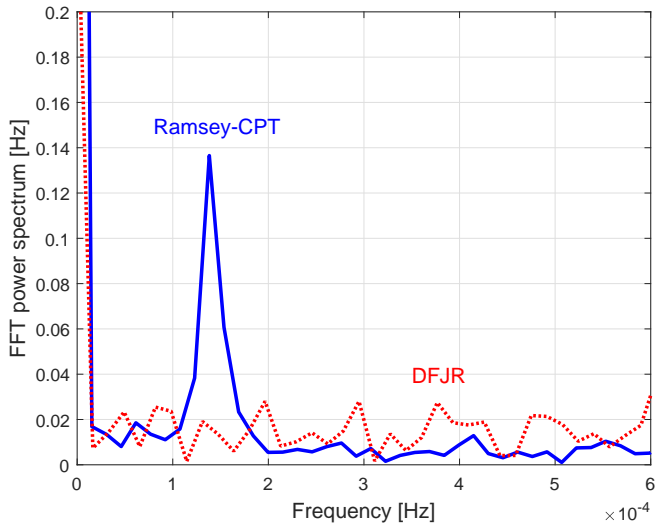}
    \caption{
    \label{image:FJR_DFJ_OPD_FFT}
    FFT power spectra of the clock frequency in Ramsey-CPT (with $T=16$ ms) and DFJR ($T_S=4$ ms, $T_L=16$ ms) methods. Sinusoidal oscillations with an amplitude of $\pm1$ MHz and about $2$ hours period are applied to the OPD of the CPT laser. An order-of magnitude reduction of the oscillations amplitude is observed with the DFJR method.   
    }
\end{figure}

We have also tested the ability of the DFJR method to mitigate light shifts induced by the CPT laser OPD. For this purpose, we applied sinusoidal oscillations to the OPD with an amplitude of $\pm 1$ MHz and a period of about $2$ hours ($\sim 1.4$Hz oscillations frequency) during clock operation. Figure \ref{image:FJR_DFJ_OPD_FFT} shows the fast Fourier transform (FFT) power spectrum of the clock frequency for the Ramsey-CPT and DFJR methods. In the Ramsey-CPT case, a strong peak is observed at the OPD oscillation frequency whereas the amplitude of this peak is reduced by an order-of-magnitude in the DFJR method. It is interesting to note that the FFT noise spectrum of the DFJR clock frequency shows an increased noise level. We attribute this degradation to the longer clock sequence in the DFJR method, as already observed in \cite{HafizPRAppl2018}.

In conclusion, we have demonstrated a simple method to reduce interrogation-related shifts in Ramsey spectroscopy and applied it to a cold-atom CPT clock. The so-called displaced frequency jumps Ramsey (DFJR) spectroscopy scheme relies on the fact that interrogation-related frequency shifts depend inversely on the dark-period and is based on consecutive Ramsey interrogations with different dark periods. Using frequency-modulation jumps, the first servo loop nulls the probing-field induced frequency shift by adding to the frequency jump a slight frequency displacement that scales with $1/T$ while the second servo loop stabilizes the local oscillator frequency. This approach presents the benefit controlling only a single physical variable (the frequency of the LO), simplifying implementation and eliminating the noise associated with controlling the LO phase. A significant reduction by more than an order-of-magnitude of the clock frequency dependence to variations of the CPT sideband ratio and of the laser one-photon detuning has been observed with the DFJR scheme, in comparison with the standard Ramsey-CPT scheme. \\

The authors acknowledge J. Elgin and C. Oates for technical help and discussions, and V. Maurice for help with the implementation of the clock control software. The Russian team was supported by the Russian Science Foundation (Project No. 16-12-10147). R. Boudot was supported by the NIST Guest Researcher fellowship and D\'el\'egation G\'en\'erale de l'Armement (DGA). This work is a contribution of NIST, an agency of the U.S. government, and is not subject to copyright.

\bibliographystyle{apsrev4-1}
\bibliography{bib_CACPT}

\end{document}